\documentclass[floatfix, reprint, amsmath, amssymb, aps, ]{revtex4-1}

\usepackage{amsmath}
\usepackage{siunitx} 
  \DeclareSIUnit\sq{\ensuremath{\Box}}
\usepackage{amssymb}
\usepackage{amsfonts}
\usepackage{amssymb}
\usepackage[pdftex]{graphicx}
\usepackage{subfigure} 
\usepackage[english]{babel}
\usepackage{braket}
\usepackage{color}
\usepackage{mathtools}
\usepackage{threeparttable}
\usepackage[colorlinks,linkcolor=blue,anchorcolor=blue,citecolor=blue,urlcolor=blue]{hyperref}
\usepackage{babel}
\makeatother

\begin{document}

\title[\texttt{achemso} demonstration]
{Enhancing the Performance of Superconducting Nanowire-Based Detectors with High-Filling Factor by Using Variable Thickness}

\author{Reza Baghdadi${^1}$}
\thanks{Both authors contributed equally to this work.}
\author{Ekkehart Schmidt${^{2,3}}$}
\thanks{Both authors contributed equally to this work.}
\author{Saman Jahani${^{4,5}}$}
\author{Ilya Charaev${^1}$}
\author{Michael G. W. M{\"u}ller${^2}$}
\author{Marco Colangelo${^1}$}
\author{Di Zhu${^1}$}
\author{Konstantin Ilin${^2}$}
\author{Alexej D. Semenov${^6}$}
\author{Zubin Jacob${^4}$}
\author{Michael Siegel${^2}$}
\author{Karl K. Berggren${^1}$}
\thanks{Corresponding author: berggren@mit.edu}

\affiliation{$^1$Research Laboratory of Electronics, Massachusetts Institute of Technology, Cambridge, MA, United States,}
\affiliation{$^2$Institute of Micro- and Nanoelectronic Systems, Karlsruhe Institute of Technology (KIT), Karlsruhe, Germany,}
\affiliation{$^3$Jet Propulsion Laboratory, California Institute of Technology, Pasadena, CA, United States,
}
\affiliation{$^4$Birck Nanotechnology Center and Purdue Quantum Center, Purdue University, West Lafayette, IN, United States,}
\affiliation{$^5$Department of Electrical Engineering, California Institute of Technology, Pasadena, CA, United States,}
\affiliation{$^6$Institute of Optical Sensor Systems, German Aerospace Center, Berlin, Germany.}

\date{\today}
\begin{abstract}
Current crowding at bends of superconducting nanowire single-photon detectors is one of the main factors limiting  the performance of meander-style detectors with large filling factors. 
In this paper, we propose a new concept to reduce influence of the current crowding effect, a so-called variable thickness SNSPD, which is composed of two regions with different thicknesses. A larger thickness of bends in comparison to the thickness of straight nanowire sections locally reduces the current density and reduces the suppression of the critical current caused by the current crowding. This allows variable thickness SNSPD to have a higher critical current, an improved detection efficiency, and decreased dark count rate in comparison with a standard uniform thickness SNSPD with an identical geometry and film quality.

\end{abstract}
\maketitle
\section{Introduction}

Superconducting nanowire single-photon detectors (SNSPD) are typically made from a thin layer of superconducting film structured into long, parallel straight nanowires that are connected in series to form meandering segments. For such a structure the switching current ($I_C$) and its local distribution have a crucial impact on detection properties and performance of the SNSPD.
The presence of localized, high-current-density regions near a right-angle bend that reduces the $I_C$ of strip conductors was theoretically predicted by Hagedorn and Hall \cite{Hagedorn.1963} in \num{1963}. The influence of the bend geometry on the switching current was investigated in theory by Clem and Berggren \cite{Clem.2011} and later, experimentally, by Henrich et al. \cite{Henrich.2012b} and Hortensius et al. \cite{Hortensius.2012}. They demonstrated that the switching current in superconducting nanowires is limited by their geometries due to current-crowding effects at corners and bends. This effect significantly limits the sensitivity of a classical meander SNSPD due to their sharp \SI{180}{\degree} turning point at the end of each straight wire section. A reduction in the switching current $I_C$ of the detector decreases the applicable bias current $I_B$ in relation to its theoretical depairing current $I_{dep}$. The drop in this ratio $\left(I_C/I_{dep}\right)$ limits the degree to which the superconducting energy gap can be suppressed in the nanowire, which defines the minimum energy of photons $\epsilon_{ph}^{min}$ that can be detected deterministically. According to the diffusion hotspot model \cite{Semenov.2005}, the proportionality of $\epsilon_{ph}^{min}$ to $I_C$ can be described as:

\begin{equation}
     \epsilon_{ph}^{min} = \frac{hc}{\lambda_c} \propto 1-\frac{I_C}{I_{dep}}.
     \label{eq:cutoff}
\end{equation}

Here, $h$ is the Planck constant and $c$ is the speed of light. The cutoff wavelength $\lambda_C$ is the smallest wavelength for which photons can be detected deterministically with an intrinsic efficiency close to unity. \\

In addition, the dark count rate is increased by current crowding: a larger current density in the bends decreases the energy barrier for vortex entry. More thermally activated vortexes, which  are the predominant cause for detector dark counts \cite{Akhlaghi.2012}, can enter the nanowire which results in an increase of the dark count rate.

Previous approaches to reduce the current-crowding effect in SNSPDs rely on the design of an optimal curvature for bends in the meander part. Optimal rounded bends require larger bending radii which in a meander decreases the filling factor, and consequently, lowers the overall detection efficiency \cite{Clem.2011}. A way to increase the bend radius without significantly compromising the filling factor is to use a spiral design \cite{Henrich.2013}. Despite convincing results of increasing the $I_C$ to $I_{dep}$ ratio from \num{0.40} to \num{0.55} \cite{Charaev.2017}, a spiral detector is not suitable for some applications: its polarization insensitivity decreases the optical coupling efficiency for polarized photons when compared to a properly aligned meander. In a detector array, any arrangement of spirals will cause blind spots and they cannot be easily integrated into photonic integrated circuits. In addition, the geometry of an optimal curvature or spiral design is more challenging to fabricate in comparison to a standard meander in the usually used electron beam lithography process.

\begin{figure}[hbt]
     \centering
     \includegraphics[width=0.95\linewidth]{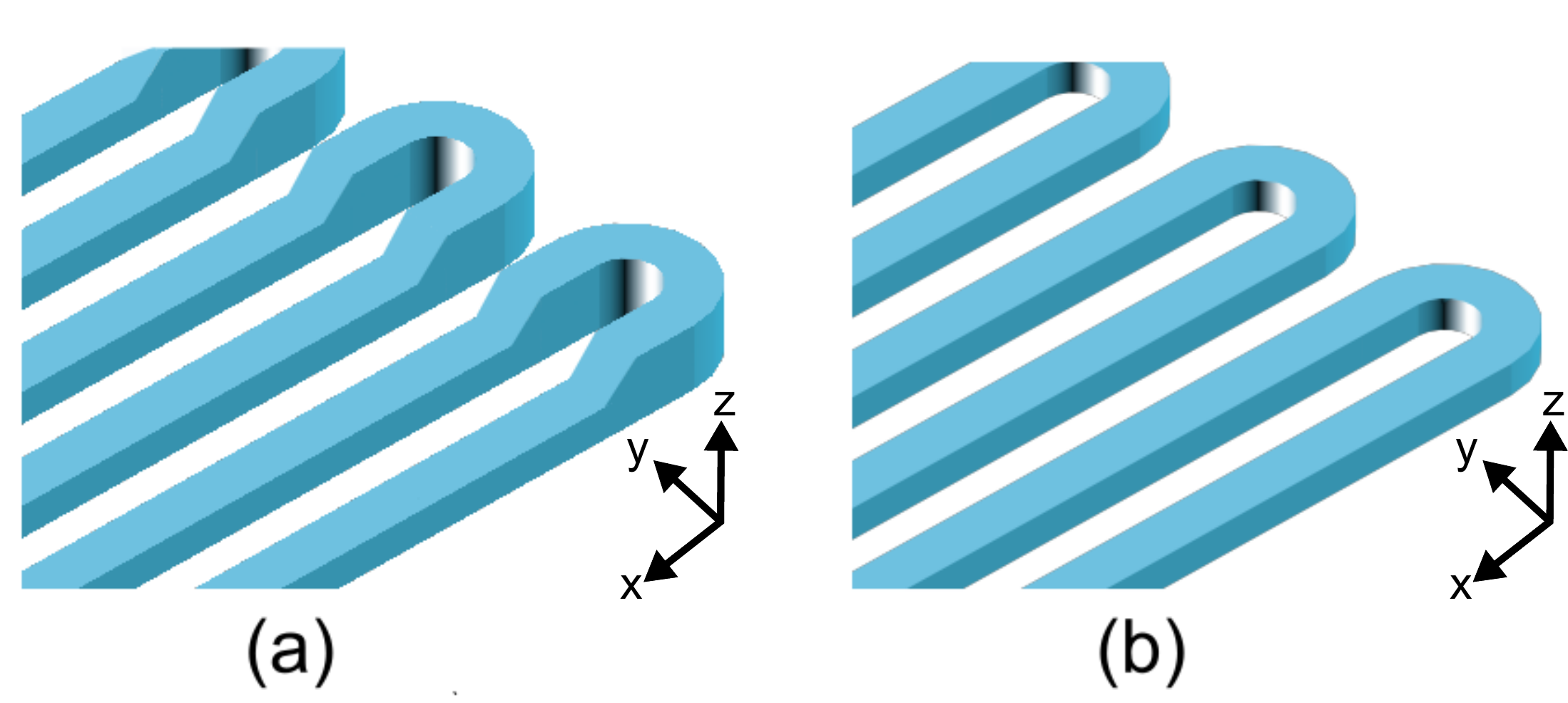}
     \caption{(a) To minimize the current-crowding effect in SNSPDs, we use different thicknesses for straight nanowire segments and bends in the superconducting film. In this design, the bends are made of a thicker film compared to the nanowire segments. (b) To understand the effect of the thicker bends, we analyzed the devices in (a) using those in (b) as controls. }
     \label{fig:fig0}
\end{figure}

As neither the optimal rounded bend approach nor the spiral approach are perfect, we propose a new approach to limit current crowding at bends.
To keep the advantages of meanders, we propose an SNSPD composed of two regions with different thicknesses. In our approach, the bends are made of thicker NbN film, while the superconducting nanowire segments (the active area of the detector) are made of thinner NbN film. Thicker bends reduce the local current density, which results in a locally increased switching current and increased vortex entry barrier. As a consequence, the switching current in the bends is higher than the switching current of the nanowires. In such a detector, which we call a variable thickness (VT) SNSPD (see Fig.\,\ref{fig:fig0}(a)), the switching current of the full detector would not be limited by the current-crowding effect at the bends. To assess the performance improvement of the VT SNSPDs, we use uniform thickness (UT) SNSPDs with a uniform thickness in straight nanowire segments and bends as controls (see Fig.\,\ref{fig:fig0}(b)). Apart from the bend's thickness, the UT SNSPDs share the same geometry.

\section{Theory}

\begin{figure}[hbt]
     \centering
    \includegraphics[width=0.99\linewidth]{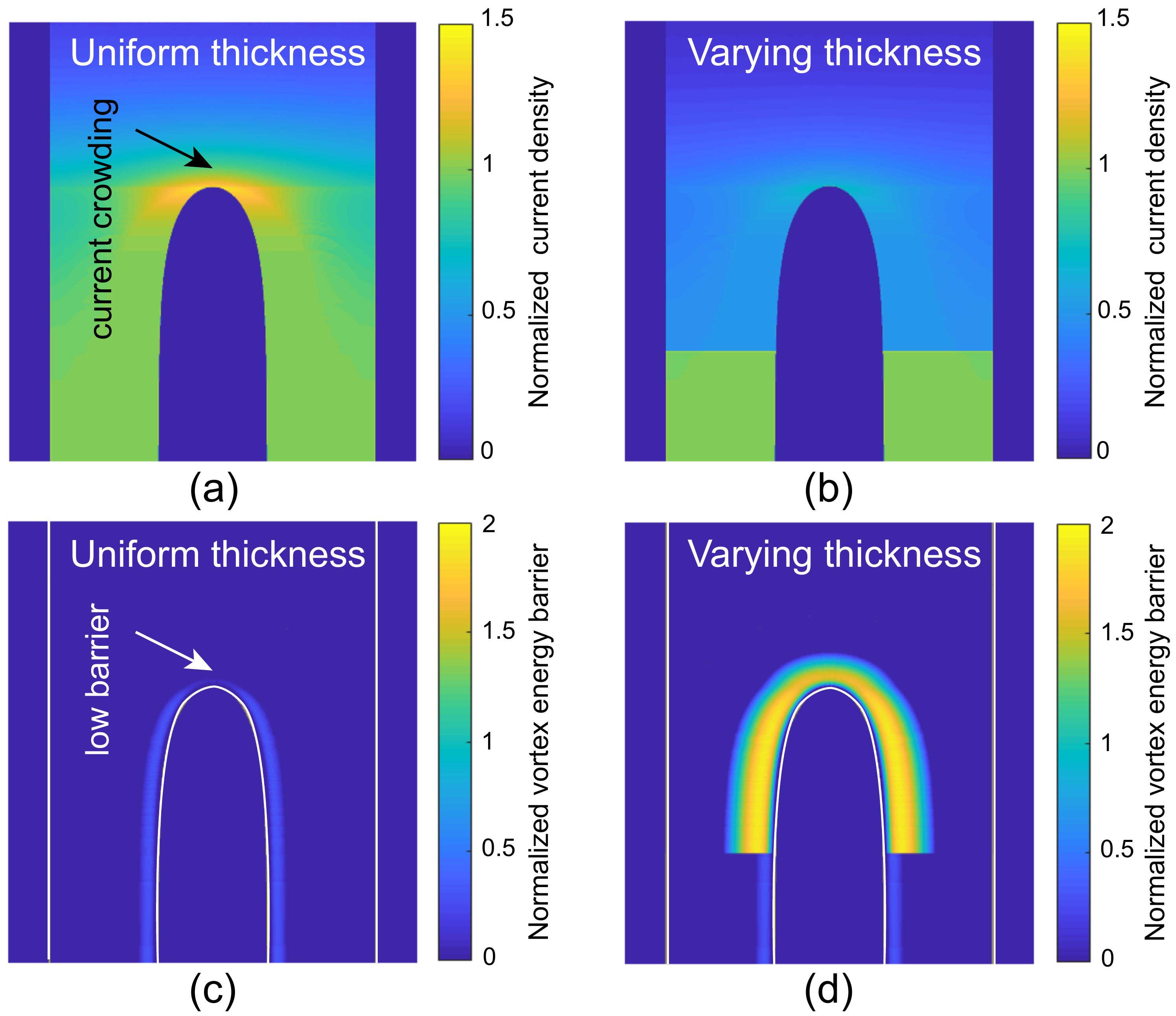}
    \caption{The bias current density in the bend normalized to the current density in the straight arm for (a) uniform and (b) variable thickness NbN SNSPDs. The width of the nanowire in the straight arm is 100~nm. The current-crowding occurs around the inner side of the curvature. Increasing the thickness near the bend reduces the current density. The vortex potential barrier normalized to the characteristic vortex energy, $\varepsilon_0$, inside the full etched region for (c) uniform and (d) variable thickness SNSPDs are also shown. The bias current is $0.7I_C^v$ in the straight section. The current-crowding weakens the vortex barrier in the uniform SNSPD. This causes a reduction in the vortex switching current and an increase in the dark-count rate. Increasing the thickness near the bend increases the self-energy of a vortex, and as a results, the switching current is enhanced to the switching current of the straight arm and the dark count rate is reduced.}
    \label{fig:fig1}
\end{figure}

To asses the potential improvement of the current distribution for the VT SNSPD and to asses the influence on the vortex entry barrier, we simulated both bend designs using a conformal mapping technique. Conformal mapping is a general approach to map a complicated 2D geometry to another 2D geometry for which an analytical solution of the current distribution exists. We use a conformal mapping technique to map a 180-degree turnaround to a half plane that has a uniform current distribution. The details of this approach can be found in Ref. \cite{Clem.2011}. Then, the inverse mapping is calculated numerically to obtain the current distribution in the actual structure.
The calculated current distribution for a turnaround with uniform thickness is shown in Fig.~\ref{fig:fig1}(a). The nanowire fill factor is $1/2$. The current-crowding effect is seen around the inner edge of the curvature. This results in a reduction in the measurable switching current, $I_C$. Increasing the nanowire thickness around the curvature helps to reduce the current density as illustrated in Fig.~\ref{fig:fig1}(b). To avoid the current crowding that potentially arises from a steep thickness step, the detector should be designed and fabricated in a way that allows a smooth transition from the thin nanowires to the thick bends.

It has recently been suggested that vortex crossing can be responsible for a phase transition in a superconducting nanowire \cite{Bartolf.2010,Bulaevskii.2011,jahani2019probabilistic}. Vortices in a thin (Type-II) superconductor are formed along the edges of the nanowire. The bias current applies a force (known as the Lorentz force \cite{tinkham_introduction_1996}) on the vortices perpendicular to the direction of the current flow, but there is a potential barrier which hinders the vortex from crossing the wire. The potential barrier of a vortex around the saddle point that is close to the inner edge of the nanowire can be approximated as \cite{tinkham_introduction_1996,Engel.2013}:
\begin{equation}
    U_v(r_v)= \varepsilon_0(r_v)\ln{\frac{2r_v}{\xi}} - \frac{\Phi_0}{c} d(r_v) \int_{0}^{r_v} j_{n}(r)dr,
    \label{eq:vortex_energy}
\end{equation}
where $\varepsilon_0={\Phi_0^2}/{8\pi^2\Lambda(r_v)}$ is the characteristic vortex energy, $\Lambda=2\lambda^2/d(r_v)$ is the Pearl length, $\lambda$ is the London penetration depth, $d(r_v)$ is the position-dependent thickness of the nanowire, $r_v$ is the distance of the vortex from the inner edge, $\Phi_0=hc/2e$ is the magnetic flux quantum, $h$ is the Planck constant, $c$ is the speed of light in vacuum, $e$ is the electron charge, $\xi$ is the coherence length, $j_{n}(r)$ is the bias current density normal to the width of the nanowire, and $r$ is the in-plane distance from the inner edge of the nanowire. The first term on the right side of Eq.~(\ref{eq:vortex_energy}) is the self energy of the vortex and the second term is the work done by the Lorentz force.

If a vortex circumvents the barrier and reaches the other edge of the nanowire and the bias current is high enough, the vortex crossing releases enough energy to cause a phase transition across the entire width of the nanowire \cite{bulaevskii2012vortex}. Increasing the bias current reduces the potential barrier. The minimum bias current that causes the barrier to vanish completely is called the vortex switching current, $I_C^v$, which is lower than the depairing current, $I_{dep}$ \cite{Bulaevskii.2011}. Even if the bias current is below the switching current and the potential barrier is not suppressed completely, vortices can be thermally excited and overcome the barrier \cite{jahani2019probabilistic}. Recent experiments have suggested that vortices overcoming the barrier is the main reason for dark counts in SNSPDs \cite{Bartolf.2010}.

Figure~\ref{fig:fig1}(c) displays the calculated vortex potential barrier based on Eq.~(\ref{eq:vortex_energy}) and the current distribution shown in Fig.~\ref{fig:fig1}(a). The current crowding near the inner edge of the bend lowers the vortex potential barrier to $0.75I_C^v$ in the straight arm of the nanowire. This causes a reduction in the switching current of the nanowire and an enhancement in the dark count rate. The current crowding can be eliminated by optimally designing the inner boundary of the curvature \cite{Clem.2011}. However, the fill factor for the optimal bend cannot be larger than $1/3$ \cite{Clem.2011}. This limits the detection efficiency of SNSPDs. 

As seen in Eq.~(\ref{eq:vortex_energy}), increasing the thickness of the nanowire increases the self energy of the vortex without changing the work done by the Lorentz force. This enhances the potential barrier within the thicker region of the nanowire as shown in Fig.~\ref{fig:fig1}(d). Hence, the switching current rises up to $I_C^v$ of the straight arm without a significant change in the detection efficiency of the SNSPD since a large portion of the SNSPD is fully etched. The higher potential barrier for vortex entry also leads to a reduction in the dark count rate. Note that Figs.~\ref{fig:fig1}(c) and \ref{fig:fig1}(d) display the potential barrier for the vortices penetrating from the inner edge into the nanowire. The potential barrier for vortices with opposite rotation which penetrate from the outer edge is even higher because the current density is lower around the outer edge of the curvature. 

\section{Design and Fabrication}

\begin{figure}[hbt]
     \centering
     \includegraphics[width=0.99\linewidth]{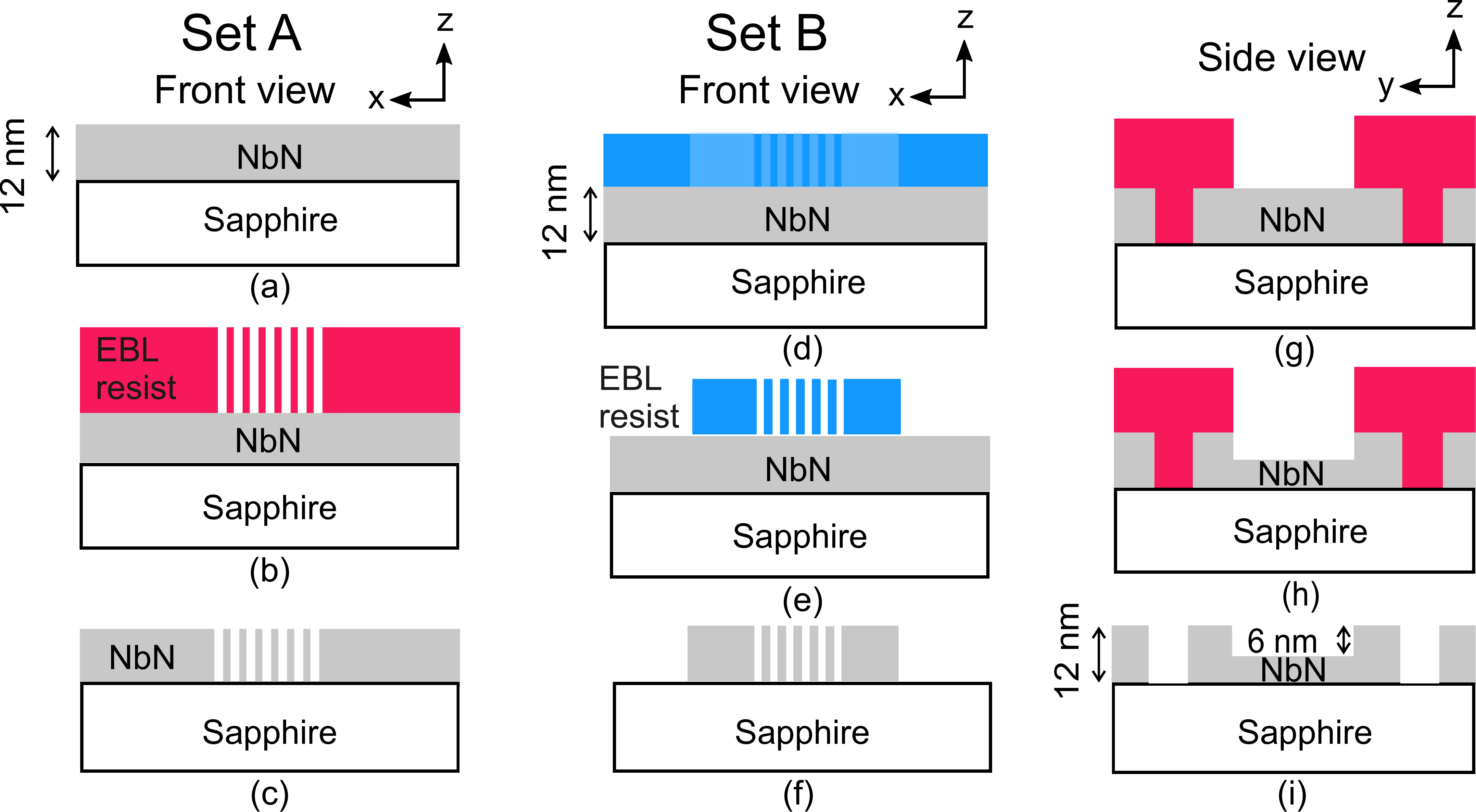}
     \caption{The fabrication sequence comprising several steps: (a) A 12 nm thick NbN film is deposited by sputtering. (b) For the SNSPD set A, a positive-tone electron beam resist (ZEP-520 A) is used to pattern the nanowires. (c) The pattern is then transferred to the NbN via CF4 reactive ion etching process. (d) For the SNSPD set B, a PMMA resist layer is used as a high-resolution negative-tone electron-beam resist. (e) The PMMA resist is exposed to a high exposure dose then developed in acetone. (f) The resist is then removed. (g) A relatively thick layer (200 nm) of positive-tone resist (set A: ZEP-520 A, set B: PMMA 950K) has been used as a hard mask. For the VT SNSPD detectors, the etch window is opened over the area which covers the active area of the detector, while for the UT SNSPD detectors, the mask opening comprises the entire detector including its bends. (h) The unprotected part of the NbN film is then slowly etched in the Ar$^+$ ion beam milling process to half of its initial thickness. (i) The resist is finally removed. For set A, the thinning was performed after, for set B prior to the nanowire fabrication.}
    \label{fig:fig2}
\end{figure}

In Fig.\,\ref{fig:fig2} the geometric design and the fabrication of detectors are depicted.
Two sets of detector chips were fabricated independently (set A and set B) in two different laboratory's. Set A was fabricated at the MIT (affiliation 1) and set B at the KIT (affiliation 2). UT  and VT SNSPDs were fabricated on each chip next to each other and underwent the same nanopatterning process. The fabrication process thereby ensured that both types of detectors have the same thickness aside from the bends.

The fabrication steps for realizing SNSPDs set A and set B SNSPDs are shown schematically in Fig.~\ref{fig:fig2}. The fabrication in both cases was performed in a three step electron-beam lithography (EBL) process. Initially, a \SI{12}{\nano\meter} thick niobium nitride (NbN) film was sputter-deposited on top of an R-plane sapphire substrate (Fig.~\ref{fig:fig2}(a)). 

For the set A SNSPDs, the meandered nanowires were patterned into NbN film in an EBL step with the use of a high-resolution positive tone electron-beam resist (ZEP 520 A) and reactive-ion etching in CF4 plasma (see Fig.~\ref{fig:fig2}\,(a)- (c)). 

The set B SNSPDs were fabricated in a slightly different way. We took advantage of the polymethyl methacrylate (PMMA) resist in its negative-tone regime as an etch mask. The PMMA is a well-known positive-tone resist, but it could serve as a high-resolution negative tone resist and experiences cross-linking when it is exposed to a very high electron-beam dose \cite{Zailer_1996}. By using EBL, the SNSPD structure was patterned into the negative-tone PMMA resist layer and then transferred to the NbN film via an Ar$^+$ milling step (see Fig.~\ref{fig:fig2}\,(d)- (f)).  

Thinning out was achieved for the full SNSPD in the UT SNSPD and for straight sections of the VT SNSPD. The chip was covered by a \SI{200}{\nano\meter}-thick positive-tone electron beam resist, which was patterned by EBL to make an opening on top of the detector. 
For the VT SNSPDs, the openings in the resist just covered the active area of the detector and protected the bends. For UT SNSPDs the entire detector area was opened. The unprotected NbN structures were etched down from \SI{12}{\nano\meter} to \SI{6}{\nano\meter} for set A and to \SI{4.3}{\nano\meter} for set B by using Ar$^+$ milling (see Fig.~\ref{fig:fig2}\,(g)- (i)). 
For set A, the etching of the active area was performed after, for set B prior to the nanowire fabrication.

For set A, we used the atomic force microscopy (AFM) technique to measure the thickness of the NbN film of the UT and VT detectors. We verified the obtained values by performing the x-ray reflectivity (XRR) of a bare NbN film before and after the Ar$^+$ion milling step. For set B, the thickness was measured using a profilometer on a witness device, which was processed in parallel to the detector sample. To have a smooth transition from the thick NbN region of the meandered region to the thinned-down region of the nanowire segments, the sample was tilted and rotated during the Ar$^+$ion milling.  

\begin{figure}[hbt]
     \centering
     \includegraphics[width=0.95\linewidth]{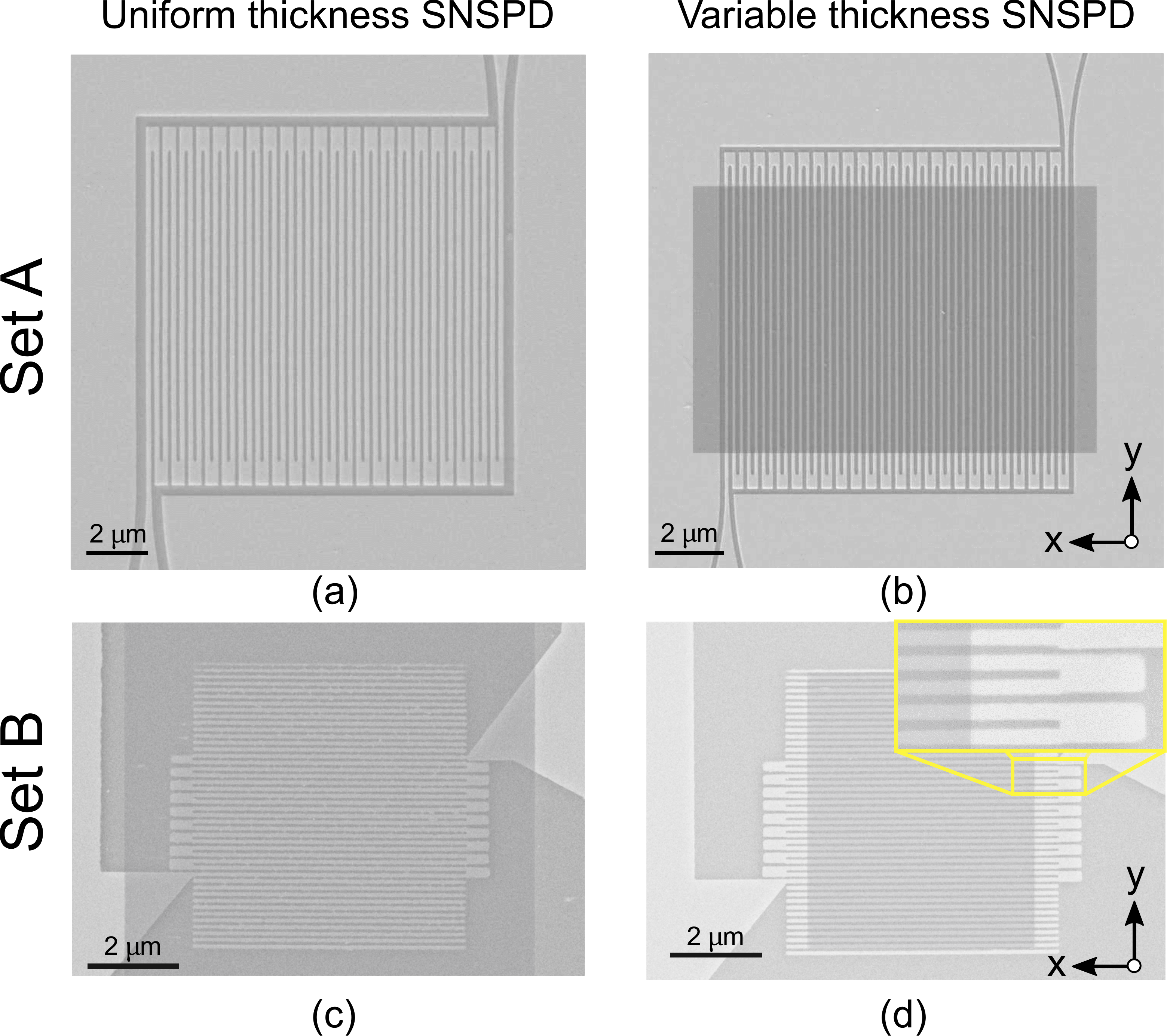}
     \caption{SEM images of two sets of meander-type SNSPDs presented in this study. In the UT SNSPD detectors (a) set A, and (c) set B, the bends and straight segments have the same NbN thickness. In  contrast, in the VT SNSPD devices of both sets A and B, (b), (d), the bends are made of a thicker NbN film than the straight nanowire segments. To realize set A detectors (a) and (b), we employed high-resolution positive tone EBL resist (ZEP520 A), and for the set B detectors (c) and (d), we used PMMA resist in its negative tone regime.}
     \label{fig:fig3}
\end{figure}

Scanning electron microscope images (SEM) of the finalized SNSPDs is shown in Fig.\,\ref{fig:fig3}. Set A SNSPDs were fabricated with nanowire widths ranging from \SIrange{100}{600}{\nano\meter}; effective areas of 9 $\mu$m$^2$, 36 $\mu$m$^2$, and 100 $\mu$m$^2$; and filling factors of \SIrange{25}{80}{\percent}. Set B SNSPDs were fabricated with \SI{85(5)}{\nano\meter} width, an active area of \SI{5x2.6}{\micro\meter} and a filling factor of \SI{60}{\percent}. The bends consisted of two turns at a $\sim$ \SI{90}{\degree} angle. To provide the best comparability, VT detectors were fabricated alongside UT detectors on the same substrate at a separation of only \SI{50}{\micro\meter}. 

It was found that the Ar$^+$ milling step could introduce degradation of superconducting properties of NbN structures in accordance with the applied etching rate:
During the fabrication of set A SNSPDs, a etching rate of \SI{2}{\angstrom\per\minute} was used. As a result of the thinning and the detector fabrication, the $T_C$ slightly decreased from \SI{9.7}{\kelvin} for $d = \SI{12}{\nano\meter}$ to \SI{9.1}{\kelvin} for $d = \SI{6}{\nano\meter}$. For the fabrication of set B SNSPDs, a higher etching rate (\SI{17}{\angstrom\per\minute}) was used, and the $T_C$ decreased from \SI{13.3}{\kelvin} for $d = \SI{12}{\nano\meter}$ to \SI{8}{\kelvin} for $d=\SI{4.3}{\nano\meter}$. To understand if the $T_C$ reduction was due to the Ar$^+$ milling process or the film's thickness, in a separate experiment we performed a direct comparison of an NbN film grown to a thickness of $d=\SI{5.5}{\nano\meter}$ and a film etched to a thickness of $d=\SI{5.5}{\nano\meter}$ with an initial thickness of $d=\SI{12}{\nano\meter}$. We applied the same sputter and etching process that we used for the fabrication of set B devices. We observed a \SI{2}{\kelvin} lower $T_C$ of the etched film when compared with the $d=\SI{5.5}{\nano\meter}$ thick film that did not undergo the etching step. 
 In summary, the higher etching rate Ar$^+$ milling step process used for set B led to a strong reduction of $T_C$, whilst the reduction of $T_C$ for set A can be explained by a reduced thickness of the film.

\section{Results}

Measuring the switching current of a VT detector and comparing it to a UT detector with similar dimensions, is an accessible and reliable way to assess the effectiveness of a thicker superconducting film at bends of a detector in order to minimize the current-crowding effect.

In the first experiment, we fabricated a set of 36 UT and VT SNSPD devices (set A) with different filling factors and then extracted the switching current of all detectors from their current-voltage curve (IVC) measurements. As mentioned earlier, all detectors were realized on the same chip and underwent the same nanopatterning process. The IVC measurements were carried out in a dipstick probe in liquid helium (LHe) with a base temperature of \SI{4.2}{\kelvin}.

\begin{figure*}[hbt]
 \centering
 \includegraphics[width=1\linewidth]{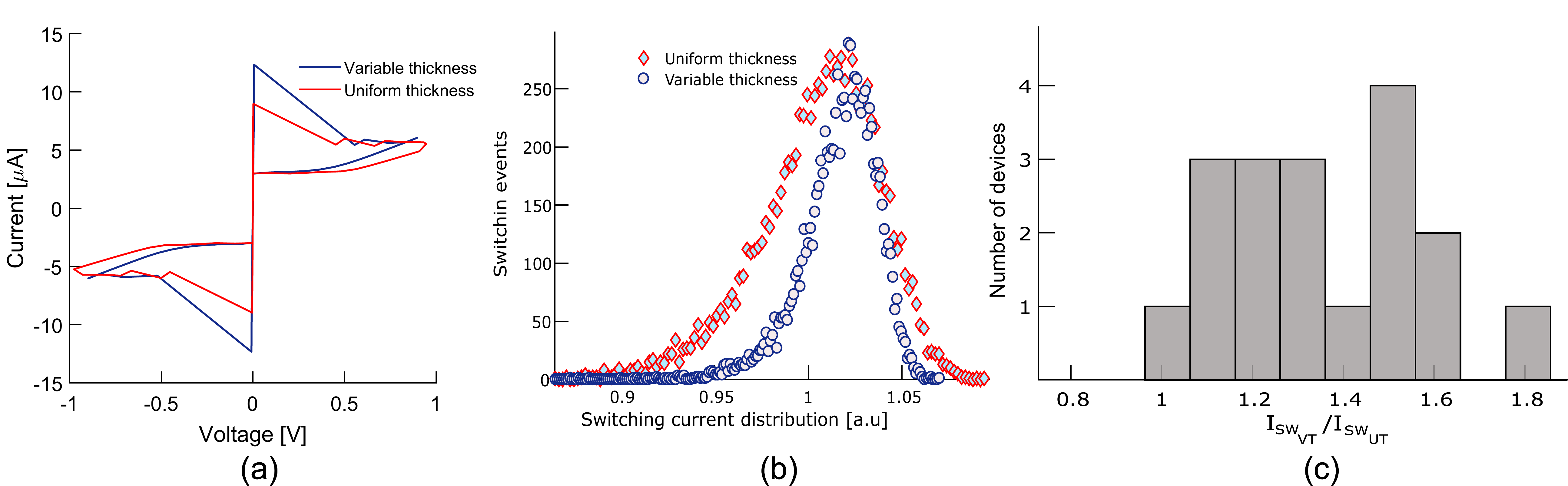}
 \caption{(a) Current-voltage characteristics at T = 4.2 K of VT (blue) and UT SNSPDs (red). The VT SNSPD shows a higher switching current. As opposed to VT SNSPD, the switching current density at bends of UT SNSPD exceeds the critical current of straight nanowire segments. (b) switching current distribution of VT (blue) and UT SNSPD  (red). The VT SNSPD possesses a narrower distribution since the energy barrier for the entry of thermally-activated vortices is higher compared to the UT SNSPD. (c) The ratio of the switching currents of VT SNSPDs compared to those of UT SNSPD.}
 \label{fig:fig4}
\end{figure*}

Figure~\ref{fig:fig4}(a) shows a typical IVC similar for both VT  and UT SNSPDs (set A). Detectors were characterized with hysteretic IVCs which is a common feature of NbN nanostructures due to low thermal conductivity, high normal sheet resistance, and high switching current values. From the IVC measurements, we extracted the switching current value of each device. As shown in Fig.~\ref{fig:fig4}(a), the VT SNSPD (set A) demonstrated a higher switching current value relative to the UT SNSPD. 
 Furthermore, we measured the histogram distribution of switching current from 10$^4$ successive IVC measurements for both UT  and VT detectors shown in Fig.~\ref{fig:fig4}(b). The VT SNSPDs exhibit a narrower distribution of switching currents over their 10$^4$ switching events. 

As our simulation results suggest, due to current crowding, the switching current density is higher at bends. Thus the energy barrier for a vortex to enter the nanowire is locally suppressed at the bends. This reduction in the energy barrier is more significant in UT SNSPDs, where current crowding is more pronounced. As a result, one expects more thermally-activated vortex crossing events in UT SNSPDs, which contributes to the broadening of the critical-current distribution.

Figure~\ref{fig:fig4}(c) summarizes the ratio between the switching current of VT SNSPD and UT SNSPD, $\left< I_{C}^{VT}/I_{C}^{UT} \right>$. Most VT SNSPDs switching current showed enhancement in the range of \SI{30}{\percent} to \SI{40}{\percent}, with a maximum value of \SI{80}{\percent}. For the set B detectors the enhancement of the switching current is comparable with an improvement by \SI{30(2)}{\percent}.

\begin{figure*}[hbt]
     \centering
     \includegraphics[width=0.6\linewidth]{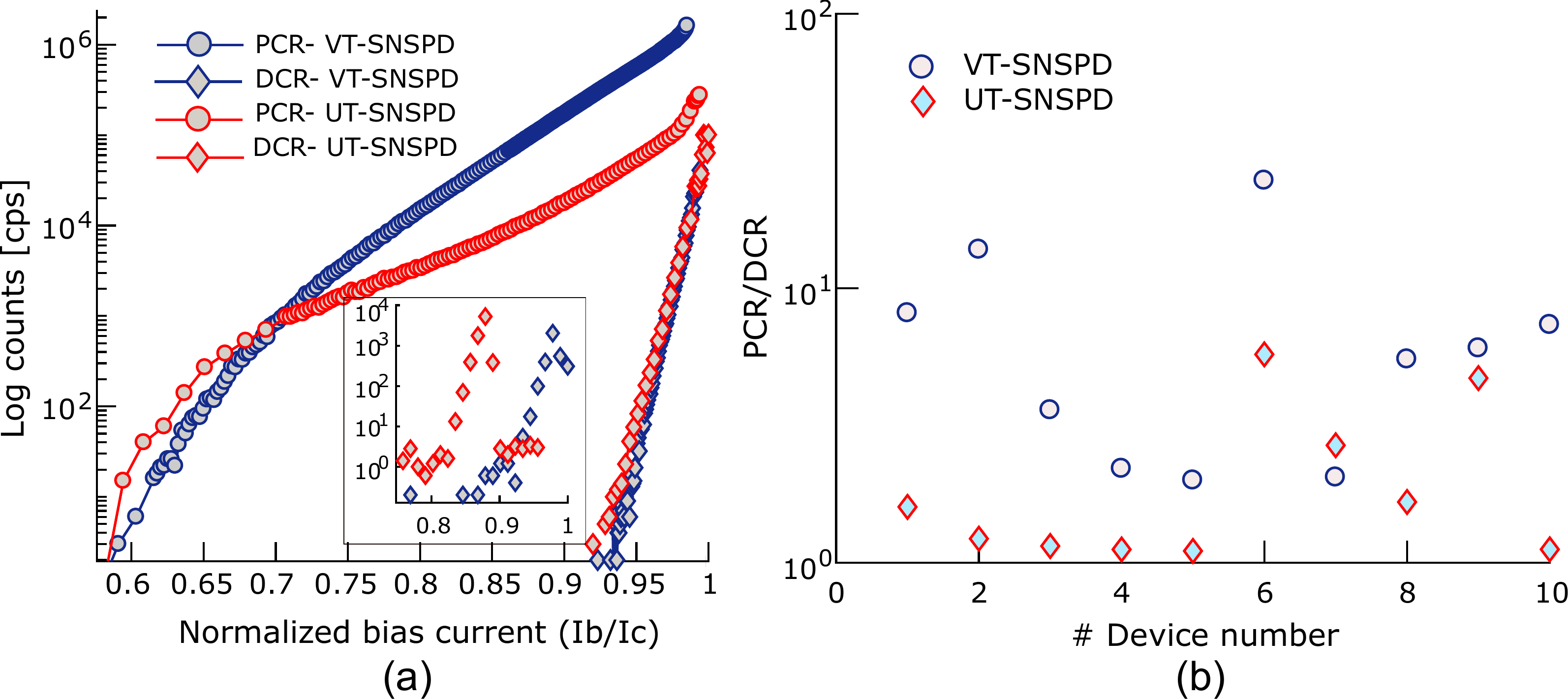}
     \caption{(a) PCR and DCR of UT  and VT SNSPDs. At high bias regime, the SNSPD with varying thickness showed a higher PCR and lower DCR  compared with the one with uniform thickness. The inset shows the DCR measurement of the same device measured in the LHe Dewar (red data) and the a cryostat with proper shielding (blue data). (b) Ratio of the detected photon count to the dark count rate is higher in SNSPDs with varying thickness. }
     \label{fig:fig5}
\end{figure*}

We also characterized the photoresponse of set A detectors. The measurements were performed inside LHe at T=\SI{4.2}{\kelvin} using \SI{1550}{\nano\meter} illumination. Figure~\ref{fig:fig5} shows a photo-count-rate (PCR), and dark-count-rate (DCR) of typical VT  and UT SNSPDs with an active area of $\SI{10}\times \SI{10}{\micro\meter}$ (\SI{100}{\nano\meter}-wide parallel nanowires with \SI{50}{\percent} filling factor) as a function of the normalized bias current. As shown, the maximum PCR of the VT detector is higher than that of the UT detector. In addition, the DCR of the VT detector is slightly lower than that of the UT detector. Since the measurements were carried out inside a LHe dewar with no proper shielding for thermally radiated photons, the DCR is relatively high. We repeated the measurements  in a closed-cycle cryostat with a better shielding for stray light and thermal radiation, and determined that the DCR of the SNSPDs reduced significantly (see inset in Fig.~\ref{fig:fig5}(a)).

We assessed several UT  and VT SNSPDs of set A, and compared the ratio of PCR to DCR $\left<PCR/DCR\right>$ of similar devices. As expected, in almost all cases but one, the $\left<PCR/DCR\right>$ ratio was higher for the VT SNSPDs (see Fig.~\ref{fig:fig5}(b)). 

\begin{figure}
     \centering
     \includegraphics[width=0.8\linewidth]{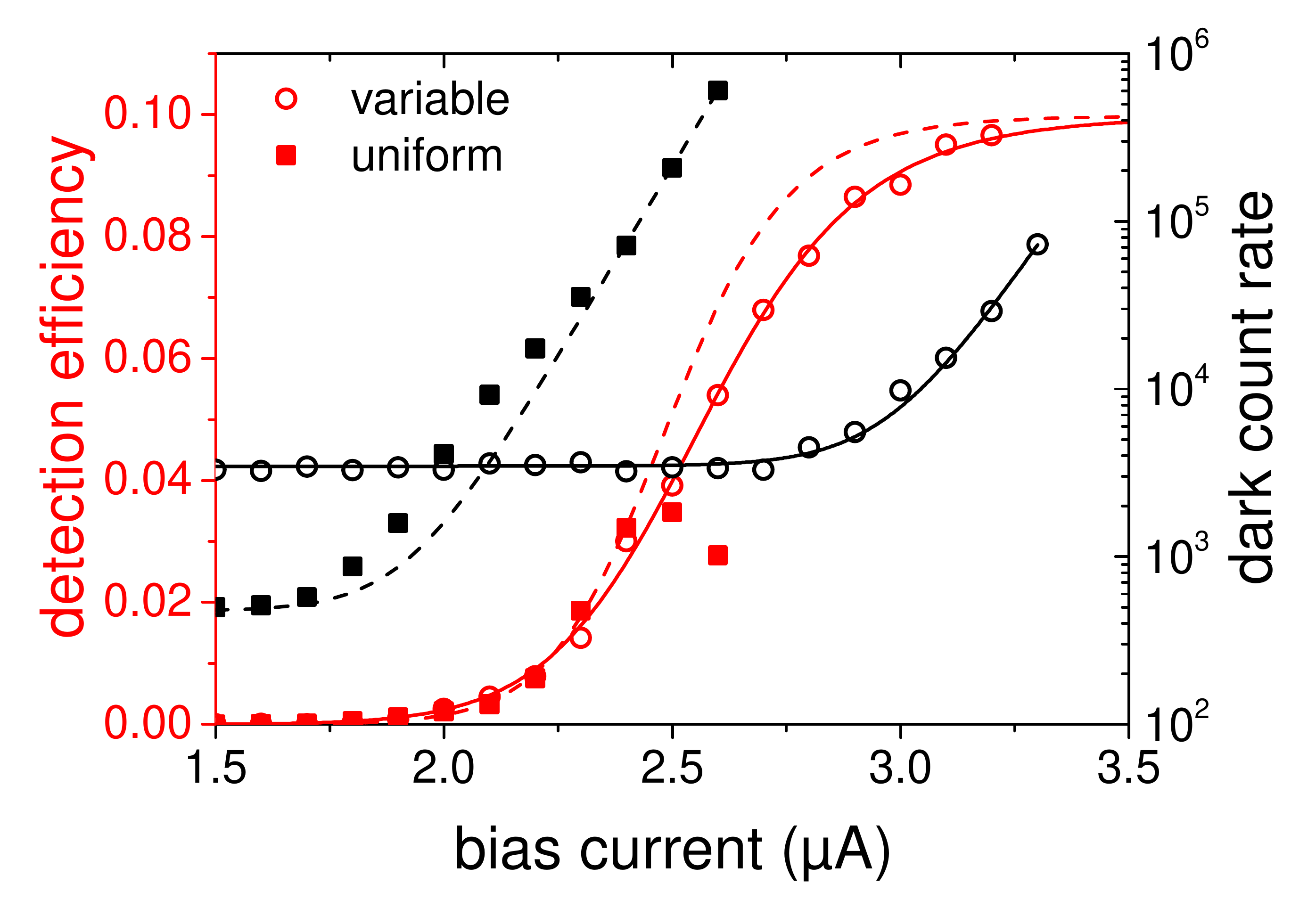}
     \caption{Bias dependence of the DE at $\lambda = \SI{500}{\nano\meter}$ for set B SNSPDs. Solid lines are fits to the data of the VT SNSPD and the dashed lines to the data of the UT SNSPD to extract the saturation and dark count level. See main text for description of the fitting process.}
     \label{fig:fig6}
\end{figure}

The detection efficiency (DE) for SNSPDs of set B with respect to the bias current is shown in Fig.~\ref{fig:fig6} for $\lambda = \SI{500}{\nano\meter}$. The DE of the detectors was calculated as $\mathrm{DE} = (\mathrm{PCR}-\mathrm{DCR})/N_{Ph}$, where $N_{Ph}$ is the number of photons incident on the active area of the detector. The method that was used to estimate $N_{Ph}$ is described in \cite{Henrich.2012}. Both detectors start to detect photons at a bias current ($I_B$) of \SI{\approx 2}{\micro\ampere}. The DE enhances with an increasing $I_B$ whilst following a sigmoidal shape for both detectors until the $I_C$ of the UT SNSPD is reached. The decrease of the calculated DE of the UT SNSPD close to $I_C$ is caused by the high dark count rate of the detector at this $I_B$ which was comparable to the photon count rate at the given point. The DE of the VT SNSPD continues to increase until it starts to level out at a DE $\approx$\SI{10}{\percent}. To extract the saturation level of the detectors, the bias dependence of DE (DE($I_B$)) was fitted with an empirical sigmodial logistic function: 

\begin{equation}
     \textrm{DE}(I_B) = \textrm{DE}_{saturation} - \frac{\textrm{DE}_{saturation}}{1+(I_B/I_{B0})^p}
\end{equation}

\noindent where $\textrm{DE}_{saturation}$ is the DE at saturation, and $p$ is the power law dependence of the DE. $\textrm{DE}_{saturation}$, $p$ and  $I_{B0}$ were used as free fitting parameters for the VT SNSPD. For the UT SNSPD, $\textrm{DE}_{saturation}$ was fixed to the saturation value extracted from the best fit to the VT SNSPD. Since both are made from the same film and have an identical geometry in their respective active areas, the DE at saturation should be the same for both detectors. In addition, for the UT SNSPD the two points measured at the highest bias currents were excluded from the fit as the dark count rate was higher than the light count rate for these points, which impacts the accuracy of the calculated DE.

The UT SNSPD reaches \SI{32}{\percent} whilst the VT SNSPD reaches \SI{97}{\percent} of the extracted saturation level. The extracted $\textrm{DE}_{saturation}$ is \SI{10}{\percent}, which is low for a saturated NbN-SNSPD. The low extrapolated DE in saturation may be explained by the small absorption length connected to the thin film in the active area and a smaller absorption efficiency of the nanowire for, with respect to the wire, orthogonal polarized photons \cite{Anant:08}.

The dark counts in Fig.~\ref{fig:fig6} were fitted with an exponential function with an offset. The offset is caused by electronic noise and is connected to a low signal-to-noise ratio (SNR) of detector pulses due to the low $I_C$ of investigated detectors in the measurement setup. Since this offset DCR is orders of magnitude lower than the true DCR at near-switching biasing, it does not affect our analysis and comparison between the two different types of devices. The exponential slope close to the respective $I_C$ revealed the intrinsic DCR of the detectors \cite{Yamashita.2011}. Due to lower $I_C$, the intrinsic DCR of the UT SNSPD is significantly higher. Furthermore, even at the same $I_B$ relative to its respective $I_C$, the intrinsic DCR of the UT SNSPD is one order of magnitude higher.

\begin{figure}
     \centering
     \includegraphics[width=0.8\linewidth]{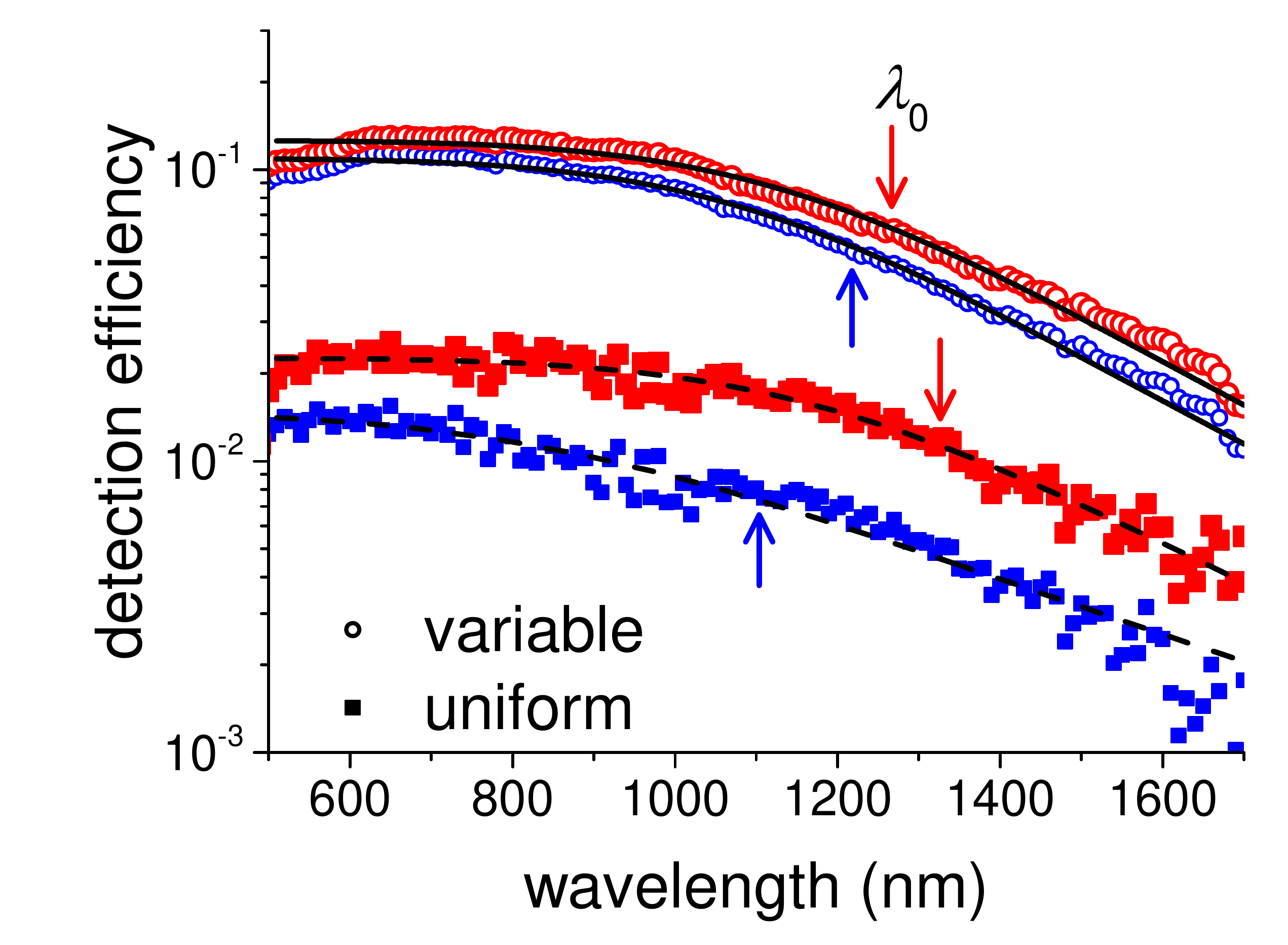}
     \caption{Spectral dependence of DE for set B VT  and UT SNSPD for $I_B = 0.85 I_C$ (blue) and  $0.9 I_C$ (red). The black lines are fits to Eq.~(\ref{eq:lambdac}).}
     \label{fig:spectralDE}
\end{figure}

The spectral dependence of the DE was measured for  VT  and UT SNSPDs (set B) at  $I_B = 0.85 I_C$ (blue) and  $0.9 I_C$ (red) for $\lambda = \SI{500}{\nano\meter}$ to $\SI{1700}{\nano\meter}$ (Fig.~\ref{fig:spectralDE}). As for the light source, a halogen lamp in combination with a monochromator was employed to select a \SI{10}{\nano\meter} wide spectral line. The use of a grating-based monochromator leads to emission of polarized light, however, the polarization was not controlled during the experiment. The light was attenuated to be in the single-photon regime, and then sent to the detector. Over the investigated spectral range, the DE was significantly higher for the VT SNSPD biased at the same relative $I_B/I_C$. For the VT SNSPD, as well as for the UT SNSPD, a plateau of the DE ($\mathrm{DE}_{plateau}$) for shorter wavelengths was seen. This plateau was followed by a roll off. The absolute DE value as well as a slight increase in DE for $\lambda = \SI{500}{\nano\meter}$ to \SI{620}{\nano\meter} for the VT detector was in good agreement with the absorption efficiency calculated in Ref.~\cite{Hofherr.2010} for polarized light orthogonal to the meander lines.
The spectral dependence of the DE can be empirically described with Eq. (\ref{eq:lambdac}) in Ref.~\cite{Henrich.2012}:

\begin{equation}
    \mathrm{DE}(\lambda) =  \mathrm{DE}_{plateau} \cdot \left( 1 + \left(\frac{\lambda}{\lambda_0}\right)^p \right)^{-1},
    \label{eq:lambdac}
\end{equation}
where $\lambda_0$ is the wavelength at which the DE reaches $0.5\mathrm{DE}_{plateau}$. The extracted $\lambda_0$ is depicted by arrows in Fig.\,\ref{fig:spectralDE} for each spectral dependence. 

Here, at a bias current of $0.9 I_C$, a larger $\lambda_0$ is observed for the UT detector in combination with a lower DE in comparison to the VT  detector. This observation is unexpected and seems contradictory which we will now try to explain. 
In the plateau, the SNSPDs are expected to operate in the deterministic detection regime with an internal detection efficiency (IDE) close to unity \cite{Ilin.2012}. As a consequence the DE should only depend on the efficiency of absorption, which is comparable as both detectors share the same film and geometry apart from bends.
While the VT SNSPD seems to operate in the deterministic detection regime which is supported by the observed bias dependence (Fig.~\ref{fig:fig6}), the low DE and the unsaturated bias dependence of the UT SNSPD convincingly show that the UT SNSPD is not in the deterministic detection regime at the applied bias current. 
Since both detectors only differ in their bend thickness, the observed effect may be explained by detection of photons in bends of the UT SNSPD: at the given bias current, the UT SNSPD may only be efficiently biased in the detector edges due to the occurrence of current crowding. In contrast, the straight parts of the nanowire are not sufficiently biased and the active area is significantly smaller which explains the low DE of the UT SNSPD in comparison to the DE of the VT SNSPD. The supposition that for a UT SNSPD the detection of low energy photons occurs in bends is supported by \cite{Semenov_Bends}. In Ref. \cite{Semenov_Bends} the detection of low energy photons in bends with occurring current crowding was shown via an asymmetry of the single-photon response in square spirals with respect to an orthogonal magnetic field. 
In addition, we have to note that $\lambda_0$ is an empirical parameter, which not necessarily equal to the deterministic cut off, that itself can depend on the wire geometry. 
In conclusion, the significant improvement of the DE on the order of a factor of ten indicates a more efficient biasing of the VT SNSPD with a dark count rate one order of magnitude lower in direct comparison to the UT SNSPD. Both detector types show a comparable $T_C$ and a coinciding bias dependence (Fig. \ref{fig:fig6}). This indicates a high comparability of both types. From the comparable design of the bends, apart from the thickness, it can be concluded that the observed improvements stem from a weakened current suppression and a larger vortex entry barrier in bends, which confirms the expected behavior.

\section{Conclusion}

We have demonstrated an effective approach to minimize the current-crowding effect in meander-type SNSPDs. We have developed a new type of SNSPD where the straight nanowire segments and bends have different thicknesses in contrast with the conventional uniform-thickness detectors. In these detectors, called VT SNSPDs, the bends are made of a thicker superconducting film which minimizes the current-crowding effect at the bends. 

In order to compare the performance of these detectors, we fabricated conventional SNSPDs in which the thickness of the superconducting film is the same for nanowire segments and bends, called UT SNSPDs. To provide a valid comparison, the VT  and UT SNSPDs were fabricated with  similar dimensions and were realized on the same chip, thus ensuring that they undergo the same fabrication processes. 

We have shown that the VT SNSPDs have a higher switching current when compared to the UT SNSPDs. This is expected, since the $I_{sw}$ of a VT SNSPD is limited to the $J_{C}$ of the nanowire segments, compered to the UT SNSPD in which the detector's $I_{sw}$ is limited to the $I_{sw}$ of the bends. 

The photoresponse measurements of the VT SNSPDs showed an increase in PCR and a decrease in DCR when compared to the UT SNSPDs. For the set B devices, the VT SNSPD showed an improved saturation level of \SI{97}{\percent} in comparison to the UT SNSPD with a saturation level of \SI{32}{\percent} at a wavelength of \SI{500}{\nano\meter} and a bias level close to their respective switching currents. As a consequence a significant increase of the detection efficiency of the VT SNSPD for the full investigated spectral range from \SI{500}{\nano\meter} to \SI{1700}{\nano\meter} is observed. 

This study shows that an increase in the bend thickness in comparison to the thickness of the active area is a promising technological approach that can be used to minimize the current-crowding effects on the bends of SNSPD devices, and it thereby allows the development of high-performance detectors.

\section*{ACKNOWLEDGMENTS}

Research was sponsored by the U.S. Army Research Office (ARO) and was accomplished under the Cooperative Agreement Number W911NF-16-2-0192. The authors also wish to thank Dr. Luqiao Liu's group at MIT for providing access to their Ar$^+$ milling tool. S. Jahani and Z. Jacob acknowledge support by DARPA DETECT. The authors like to thank Marco Turchetti, Navid Abedzadeh, Brenden Butters, and Ashley Qu for comments on the manuscript.

\bibliography{references}
\end{document}